\documentclass{article}

\usepackage{algorithm}
\usepackage{algpseudocode}
\usepackage{enumitem}
\usepackage{tabu}
\usepackage{xcolor}
\usepackage{cite}
\usepackage{hyperref}
\usepackage{amsmath,amssymb,amsfonts}
\usepackage{graphicx}
\usepackage{textcomp}
\usepackage{url}

\newcommand{\orcid}[1]{\href{https://orcid.org/#1}{\includegraphics[scale=0.16]{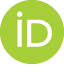}}}
\providecommand{\keywords}[1]
{
  \textbf{ \small	\textit{Keywords---}} #1
}

\begin{document}

\title{Flexible remote attestation of pre-SNP SEV VMs using SGX enclaves}

\author{Pedro Antonino \orcid{0000-0002-5627-0910} $^{1}$, Ante Derek \orcid{0000-0002-2437-8134} $^{2}$, \\ 
and Wojciech Aleksander Wo\l{}oszyn \orcid{0000-0002-5725-1658} $^{1,3,4}$ \\
\small $^{1}$ The Blockhouse Technology Limited, Oxford, UK \\
\small $^{2}$ Faculty of Electrical Engineering and Computing, University of Zagreb, Zagreb, Croatia \\
\small $^{3}$ Mathematical Institute, University of Oxford, Oxford, UK \\
\small $^{4}$ St Hilda's College, Oxford, UK}

%%% Shortcuts for notation 
\newcommand{\IntelRotk}{IntelLtk}
\newcommand{\AmdRotk}{AmdLtk}
\newcommand{\Qek}{Qek}
\newcommand{\PspSn}{PspSn}

\newcommand{\E}{\textit{E}}
\newcommand{\QE}{\textit{QE}}
\newcommand{\IntelRot}{\textit{Intel RoT}}
\newcommand{\RP}{\textit{RP}}

\newcommand{\PSP}{\textit{SP}}
\newcommand{\AmdRot}{\textit{AMD RoT}}
\newcommand{\SVM}{\textit{SVM}}
\newcommand{\GO}{\textit{GO}}

\newcommand{\TO}{\textit{TO}}

\DeclareRobustCommand{\xjoinrel}{\mathrel{\mkern-3.5mu}}
\newcommand{\ltr}[3]{\mathrel{[#1]}\xjoinrel\mathrel{-}\xjoinrel\mathrel{[#2]}\xjoinrel\rightarrow\xjoinrel\mathrel{[#3]}}

\date{}

\maketitle

\begin{abstract}
  We propose a protocol that explores a synergy between two TEE implementations: it brings SGX-like remote attestation to SEV VMs. We use the notion of a \emph{trusted guest owner}, implemented as an SGX enclave, to deploy, attest, and provision a SEV VM. This machine can, in turn, rely on the trusted owner to generate SGX-like attestation proofs on its behalf. Our protocol combines the application portability of SEV with the flexible remote attestation of SGX. We formalise our protocol and prove that it achieves the intended guarantees using the Tamarin prover. Moreover, we develop an implementation for our trusted guest owner together with example SEV machines, and put those together to demonstrate how our protocol can be used in practice; we use this implementation to evaluate our protocol in the context of creating \emph{accountable machine-learning models}. We also discuss how our protocol can be extended to provide a simple remote attestation mechanism for a heterogeneous infrastructure of trusted components.  
\end{abstract}


\section{Introduction}

% Trusted Execution Environments intro: properties and benefits
Primitives to implement a Trusted Execution Environment (TEE)~\cite{Maene18} are becoming a common feature of modern processors. Such an environment typically allows a program to execute confidentially whereby not even the operator can tell what instructions and data are being used, we refer generically to such a protected execution as an \emph{isolated computation}. Intel's Software Guard Extensions (SGX)~\cite{Costan16,SGXSDKRef}, AMD's Secure Encrypted Virtualization (SEV)~\cite{sev,sevref}, and  ARM's TrustZone~\cite{Pinto19} are examples of TEE implementations available. They are designed to address different application scenarios, but they all share similar core capabilities.

% Competing architectures SGX vs SEV
Intel's SGX and AMD's SEV provide competing TEE architectures that isolate computations at different levels of granularity. While SGX was designed to isolate (part of) an operating system process (an \emph{enclave} in SGX terminology), SEV isolates an entire virtual machine (VM). Given these design choices, SGX does not offer the same level of application portability that SEV does. An application has to be redesigned to be made SGX-aware, whereas SEV allows it to be seamlessly executed within a confidential machine. This portability comes at the price of a having a typically larger \emph{trusted computing base}. While SGX allows developers to finely tune which functions and data are part of the enclave, SEV VM would usually contain an entire operating system (OS) together with the relevant applications to be executed. The larger the trusted computing base, the more prone to bugs and vulnerabilities it is.

% Remote attestation
\emph{Remote attestation} is the process that establishes trust on an isolated computation. It consists of a protocol that produces evidence that a given computation has been properly isolated and, typically, provides a way to establish a secure channel with the isolated computation. While SGX provides a very flexible mechanism to attest enclaves, SEV (pre-SNP\footnote{We call \emph{SEV pre-SNP} the SEV implementations predating SEV SNP (Secure Nested Paging)~\cite{sevsnp20}, i.e., the original SEV implementation~\cite{sev} and SEV-ES (Encrypted State)~\cite{seves}.}) relies on a very restrictive scheme for that. 
While SGX's attestation is \emph{undirected}, namely, any third-party can establish trust on a given enclave, SEV proposes a mechanism by which only a designated party, called the \emph{guest owner}, can meaningfully attest (and provision) its SEV VM.

We propose, formalise, verify, implement and evaluate a new protocol that provides \emph{SGX-like remote attestation to a SEV VM}. Broadly speaking, it relies on a special enclave that we design, the \emph{trusted guest owner}, that is responsible for deploying, attesting, and provisioning the SEV VM it owns. Moreover, while operating, this VM can request the generation of attestation reports, on its behalf, to the trusted guest owner --- in the similar way to how an enclave can create an attestation report in the SGX architecture. Our innovative combination of TEE implementations brings together the best of both worlds, namely, the application portability of SEV and the flexible attestation of SGX.  However, our protocol requires two separate platforms: a SGX-capable machine to run the trusted guest owner and a SEV-capable one for the confidential VM. Therefore, the flexibility comes at a price of a larger trusted computing base.

A composition of systems  does not necessarily yield a scheme that inherit the security properties of the components --- for instance, composing secure protocols does not automatically yield a secure scheme. Finding a protocol design that ensures the desired attestation properties was therefore challenging, and that is also why we formally analyse our protocol.
We use the Tamarin prover~\cite{Meier2013} to model our protocol and to verify that it indeed achieves the desired goal of authenticity and integrity of attestation proofs. Additionally, we verify security properties of SGX and SEV attestation as used in our protocol --- the authenticity of the SGX attestation proofs and secrecy of SEV provisioned secrets, respectively. All results hold in a general setting with unbounded number of participants and sessions, assuming a Dolev-Yao attacker~\cite{Dolev83} and a fine-grained threat model that, for example, allows the attacker to run enclaves of its choice alongside the trusted guest owner and compromise some TEE platforms. 

To demonstrate the protocol, we implement the protocol participants --- namely the trusted guest owner, the SEV guest VM attestation library and several sample SEV guest VMs. Furthermore, we evaluate our protocol by harnessing it to implement a notion of accountability for machine learning models --- i.e. creating a cryptographic report that ties a model to the technique and data used to generate it. Our evaluation demonstrates that our protocol incurs a negligible overhead while delivering on its security promises.

Some recent TEE implementations such as SEV SNP (Secure Nested Paging)~\cite{sevsnp20} and Intel's TDX (Trust Domain eXtensions)~\cite{TDX} were designed to provide a combination of remote attestation flexibility and application portability that is similar to what our protocol achieves with the proposed pairing of SGX and SEV.
However, these technologies are still not widely available and the underlying attestation mechanisms and primitives have not yet been fully scrutinized by the research community.
% However, these new implementations are still fairly immature.
%; for instance, a feature that allows measured boot available in the AMD-provided firmware for SEV pre-SNP VMs is not available for the SEV SNP firmware. 
Since Q1 2023, there a limited number of Intel CPU models supporting TDX available on the market~\cite{google_tdx}. However, at the time of writing (May 2023) the general availability of TDX remains planned for future Indel Xeon family releases and no major cloud provider offers TDX capable CPUs. Hardware support for SEV SNP was launched two years ago (Q2 2021), but software support is  somewhat lagging and SNP patches were being merged to Linux kernel in Q3 2022. While some cloud provider do offer SEV SNP enabled hardware, we found that no major provider exposes the flexible attestation interface to the end user. Microsoft Azure, for example, only allows their pre-approved VMs to be launched as SEV SNP guests, and exposes attestation only through Azure-issued JWT (JSON Web Token) tokens~\cite{azure_sev_snp}. Our protocol, on the other hand, is based upon TEE implementations that are reasonably mature and have been available for quite a few years. Even when these new technologies catch up, our protocol will still be relevant for platforms, legacy or not, that do not support SEV SNP or TDX but support SEV pre-SNP.

%% Investigation on the interplay between TEE implementations and how to combine them to create a whole that is better than the parts.
Our protocol sheds light in a new line of research, that is, finding synergies between TEE implementations. In our case, we create a protocol that brings together a pairing of a SGX enclave and a SEV VM in a way that it offers better features than both elements individually. Moreover, it can be extended to handle a related problem, namely, how to attest a homogeneous infrastructure of trusted components. Our protocol can be seen as a degenerate case of this problem where the trusted guest owner deploys a simple trusted infrastructure consists of a single SEV VM. However, our ideas could be carried over to the context of a generic \emph{trusted deployer} that could deploy, attest and provision a complex composition of trusted components.

% Summary of contributions with bullet points.
We sum up our contributions in the following:
\begin{itemize}
    \item We propose a protocol that brings SGX-like remote attestation to SEV VMs, creating a synergy that combines the application portability of SEV with the flexible remote attestation of SGX.
    \item We formalise our protocol and verify it achieves the desired guarantees/goals using the Tamaring prover.
    \item We created implementations for our trusted owner and several protocol-compatible SEV VMs. \footnote{We make the protocol implementation, the sample systems used for evaluation, as well as the formal model and proofs publicly available~\cite{repo} under a permissive open source license.}
    \item We carried out an evaluation that demonstrates how our protocol can be used to implement a notion of accountability for machine learning models. It also shows that it delivers its guarantees with negligible overhead. 
    \item The proposal of our protocol sheds light in a new line of research consisting of exploring synergies between different TEE implementations.
    \item We discuss how our protocol can be extended to provide a simple way to remotely attest an infrastructure involving heterogeneous trusted components.
\end{itemize}

% Outline
\noindent
\textit{Outline.} In Section~\ref{sec:background}, we introduce relevant background. Section~\ref{sec:flexible_attestation} introduces our protocol, together with minimalist and abstract versions of SEV and SGX attestation protocols, presents the formalisation of our protocol and discuss the properties that we were able to verify using Tamarin, and demonstrate an application of our protocol together with an evaluation of how it fares in practice. Section~\ref{sec:related_work} discusses some of the works related to ours, whereas in Section~\ref{sec:conclusion}, we present our concluding remarks.

\section{Background}
\label{sec:background}

In this section, we introduce the background elements that are necessary for understanding the rest of our paper.

\subsection{SGX}

\begin{figure}[H]
  \centering
  \includegraphics[width=.5\columnwidth]{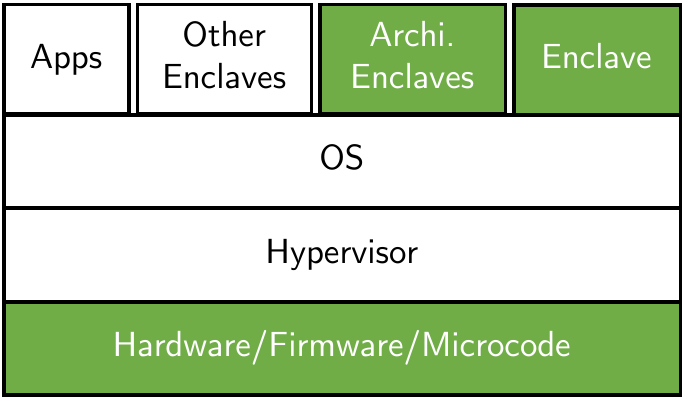}
  \caption{SGX enclave trusted computing base in green.}
  \label{fig:sgx_tcb}
\end{figure}

% What is SGX up? Focus on attestation primitives.
Intel's SGX (Software Guard eXtensions)~\cite{Costan16} allows an untrusted host process to create a protected virtual-memory range where integrity-protected and confidential code and data are hosted; this protected area is called an \emph{enclave}. SGX extends Intel's traditional instruction set with privileged instructions to create, initialise, and dispose of this protected memory range and also to non-privileged instructions to execute enclave code~\cite{SGXSDM}. A number of hardware and software components take part in enforcing the integrity and confidentiality of an enclave's execution and in attesting these properties. These elements together with the enclave code itself form the \emph{trusted computing base} (TCB) of that enclave, which is depicted in Figure~\ref{fig:sgx_tcb}; green elements are trusted, the others are not. At the lowest level, we have the trusted SGX hardware, comprising CPU package and Memory Encryption Engine~\cite{Gueron16}, and low-level code; they ensure the integrity, confidentiality and freshness of the enclave's protected memory area. Privileged code is \emph{untrusted}: privileged instructions cannot be executed in enclave mode. Hence, an enclave has to delegate to untrusted code, in the form of the OS/hypervisor, the execution of system calls, for instance. An enclave does not automatically trust other enclaves; they are isolated from one another. There are, however, some especial \emph{architectural enclaves} which are trusted. They play a fundamental part in the \emph{attestation process}, namely, in the protocol by which an enclave provides to a counterpart evidence that it is indeed a valid isolated computation executing on an authentic platform. This process attests, in fact, the entire TCB: it provides the \emph{digest} (or \emph{measurement}) of the code loaded into the enclave, and information about the version of the architectural enclaves used and the SGX hardware and low-level code. We elaborate on this process/protocol later. Applications in user-space are also not trusted by the enclave. We refer generically to the untrusted components around an enclave in a SGX platform as the \emph{SGX host}.

\subsection{SEV}

\begin{figure}[H]
\centering
  \includegraphics[width=.4\columnwidth]{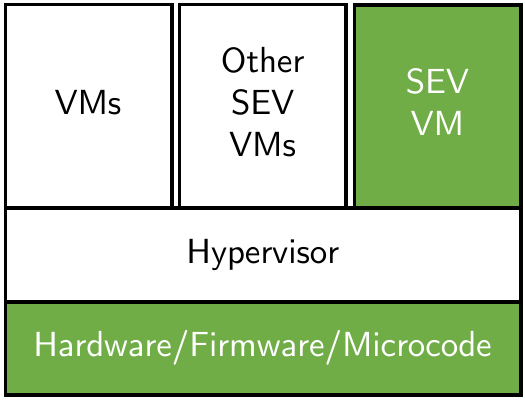}
    \caption{SEV VM trusted computing base in green.}
  \label{fig:sev_tcb}
\end{figure}

% SEV high-level; should we mention TCB?
AMD's SEV (Secure Encrypted Virtualization)~\cite{sev,sevsnp20} proposes an architecture to support \emph{confidential} virtual machines (VMs), which we refer to as \emph{SEV (guest) VMs}. This TEE implementation was designed so that even if the host (hypervisor included) is untrusted, it is unable to peek into the execution of a SEV guest VM. As for SGX, the AMD's typical instructions set was extended to incorporate directives to manage SEV VMs~\cite{sevref}. The TCB of a SEV guest machine is illustrated in Figure~\ref{fig:sev_tcb}. It consists of its own code plus SEV hardware and firmware, especially in the form of the Secure Processor - also known as Platform Security Processor, or PSP. Note that other SEV VMs are not trusted; they are isolated from one another. Other non-SEV VMs are untrusted as well. Similarly to what we do for enclaves, we refer, generically, to the untrusted elements surrounding a SEV VM in a platform as the \emph{SEV host}.

The SEV architecture has evolved from (original) SEV~\cite{sev}, to SEV-ES (Encrypted State)~\cite{seves}, and recently to SEV-SNP (Secure Nested Paging)~\cite{sevsnp20}. SEV-ES brings extra confidentiality guarantees when a switch from an trusted to an untrusted execution takes place, namely, the contents of the registers storing the state of the confidential VM are protected/encrypted before the switch occurs. SEV-SNP brings integrity guarantees that are not offered by the former two SEV versions. It also brings a form remote attestation that is more flexible than SEV and SEV-ES. We discuss an abstract version of the pre-SNP attestation protocol later.

The difference in the level of granularity for the isolated computations between SGX and SEV has relevant practical consequences. In SGX, a simple (part of a) process is isolated, as opposed to an entire VM in SEV. Therefore, the TCB for a SEV isolated computation tends to be much larger than that of a SGX computation, making it potentially more vulnerable to bugs and design flaws. However, the fact that an entire OS (and its priviledged instructions) is part of the trusted world makes this architecture more attractive in terms of application portability. An application that was not designed specifically to target a SEV VM can seamlessly (i.e. without modification) execute inside one. The same cannot be said of SGX: typically, applications have to be significantly redesigned to fit their enclave model.

\subsection{Tamarin prover}

Tamarin prover~\cite{Meier2013} is a tool for modeling security protocols and reasoning about their properties in the symbolic model of cryptography. Protocols are specified using \emph{multiset rewriting rules}, while the security properties are specified either as guarded first-order logic formulas over execution traces or as observational equivalences. Proofs can be carried manually using the interactive mode or in a automated fashion where the procedure can be further tuned by supplying a \emph{proof oracle} that prioritises available proof steps. 

Tamarin prover has been successfully used to analyse, discover vulnerabilities and provide machine-verifiable proofs of various security properties for real-world protocols such as TLS v1.3~\cite{Cremers2017TLS}, smartcard payment protocols~\cite{Basin2021EMV}, 5G authentication protocols~\cite{Basin5G2018}, and many others. In the area of trusted hardware, the tool has been used for analysis of a Direct Anonymous Attestation protocol based on the Trusted Platform Module (TPM) technology~\cite{WhitefieldDAA2019,Wesemeyer20}. 

\section{Flexible SEV pre-SNP remote attestation using SGX}
\label{sec:flexible_attestation}

In this section, we introduce a protocol that combines SGX and SEV attestation protocols in a way that it enables the flexible attestation of SEV machines. 
% The pre-SNP SEV protocol is too strict in the sense that it produces an attestation proof and provisioning channel that is \emph{targeted}, namely, only a specific party (the guest owner) can remotely attest and provision the machine; our protocol relax this requirement.
We begin by describing abstract versions of SGX and SEV attestation protocols, which we later combine to create our flexible SEV attestation protocol. We formalise these concepts using Tamarin and use this prover to verify that our protocol gives the desired security guarantees. Moreover, we present a concrete implementation (and execution) of our protocol, and close this section with a discussion on some interesting extensions to our protocol and its limitations. In this paper, we assume that side-channels attacks are possible and that the attacker can corrupt and extract secrets from arbitrary SGX/SEV platforms, enclaves, and VMs, except for the \emph{specific} platforms, enclaves, and VMs used in the protocol sessions. We claim (and formaly verify) that the proposed protocol provides a level of robustness to those attacks.

\subsection{Remote attestation for SGX enclaves}
\label{sec:sgx_attestation}

% What is SGX up? Focus on attestation primitives.
Intel has proposed two mechanisms to perform the remote attestation of an enclave: Enhanced Privacy ID (EPID)~\cite{Johnson16,Knauth18} and Data Center Attestation Primitives (DCAP)~\cite{Scarlata18}. We present a minimalist protocol for remote attestation inspired by DCAP but that abstracts away its complexity and details, focusing on its broad trust guarantees and functionality. It should be straightforward to adapt our protocol to work with the fully-fledged DCAP or EPID.

Our SGX attestation protocol involves four parties: the attested enclave \E{}, the quoting enclave of the attested platform \QE{}, Intel's Root of Trust service \IntelRot{}, and a relying party \RP{}. Broadly speaking, \QE{} is a trusted architectural enclave that runs in the same platform as \E{} and is certified by \IntelRot{}, and it creates proofs to attest \E{} to \RP{}. Note that our italicised notation here denotes the \emph{name} of the participants in our protocol. So, \QE{} is not an abbreviation for quoting enclave in general but an \emph{identifier} denoting the attested quoting enclave that participates in our protocol. We adopt this notation consistently for the participants involved in the protocols that we describe in this paper.

\subsubsection*{Protocol goal.} The protocol produces an attestation proof for \E{} consisting of a \emph{quote} in SGX terminology and a SGX platform certificate. It authenticates \E{}'s TCB. The platform certificate also contains the Platform Provisioning ID (PPID) uniquely identifying the platform instance. The quote also contains a piece of data \textit{D} that is provided by \E{}. Any relying party can, then, cryptographically validate this proof and be convinced that this quote was generated on a platform identified by PPID using the given TCB and that \E{} provided \textit{D} when the protocol was executed. 

\subsubsection*{Threat model and trust assumptions.} We assume that the platform in which \E{} is deployed has not been compromised but the attacker controls the SGX host, i.e. untrusted platform elements, and the network. So, it can arbitrarily influence communications and computations executed by these elements, and create other enclave instances. The attacker has access to compromised SGX platforms to which is can deploy enclaves. A compromised platform would allow the attacker to have access to the cryptographic keys managed by the quoting enclave and, hence, to construct arbitrary quotes that validate as correct quotes from that particular platorm. The enclave itself is known, and the attacker can deploy it at will on any platform of its choice. However, the entire attested TCB, including \E{} and \QE{}, and \IntelRot{} are trusted. Hence, the attacker can only interact with them in the ways prescribed by their implementation. We assume that the attacker cannot perform fork attacks or rollback attacks on our enclave. This is a reasonable assumption since the enclaves state will be entirely in-memory with no persisted data.

\subsubsection*{Cryptographic schemes.} Our protocol relies on the following cryptographic schemes:
\begin{itemize}
    \item \IntelRot{} uses an asymmetric signature scheme with key-pair generation function $agen_{IR}()$, signing function $asign_{IR}(m,k)$, and verification function $averi_{IR}(m,s,k_{pb})$, $m$ is a message, $s$ is a signature, $k$ is a private key, and $k_{pb}$ a public one. We use the same notation with a similar meaning when defining other asymmetric signature schemes;
    \item \IntelRot{}'s long term key pair $(\IntelRotk_{pb},\IntelRotk)$, public and private elements, respectively, is generated using $agen_{IR}()$ and used by it to issue SGX platform certificates;
    \item \QE{} uses the asymmetric signature scheme with functions $agen_{QE}()$, $asign_{QE}(m,k)$ and $averi_{QE}(m,s,k_{pb})$;
    \item \QE{} key pair $(\Qek_{pb},\Qek)$, public and private elements, respectively, is generated using $agen()_{QE}$ and used by the quoting enclave to issue attestation quotes.
\end{itemize}

We assume throughout the paper that all cryptographic payloads are tagged with labels describing the payload structure and intent of the message. For example, the payload in the certificate $C_{QE}$ below is $\langle{}\hbox{'\texttt{sgx\_platform\_certificate}'}, \Qek_{pb}, ppid\rangle{}$. However, we leave the type tags out of the protocol description to simplify notation. Of course, we include the tags in the formal model and in the protocol implementation.

\subsubsection*{Protocol.} 
\begin{figure*}
    \centering
    \includegraphics[width=\textwidth]{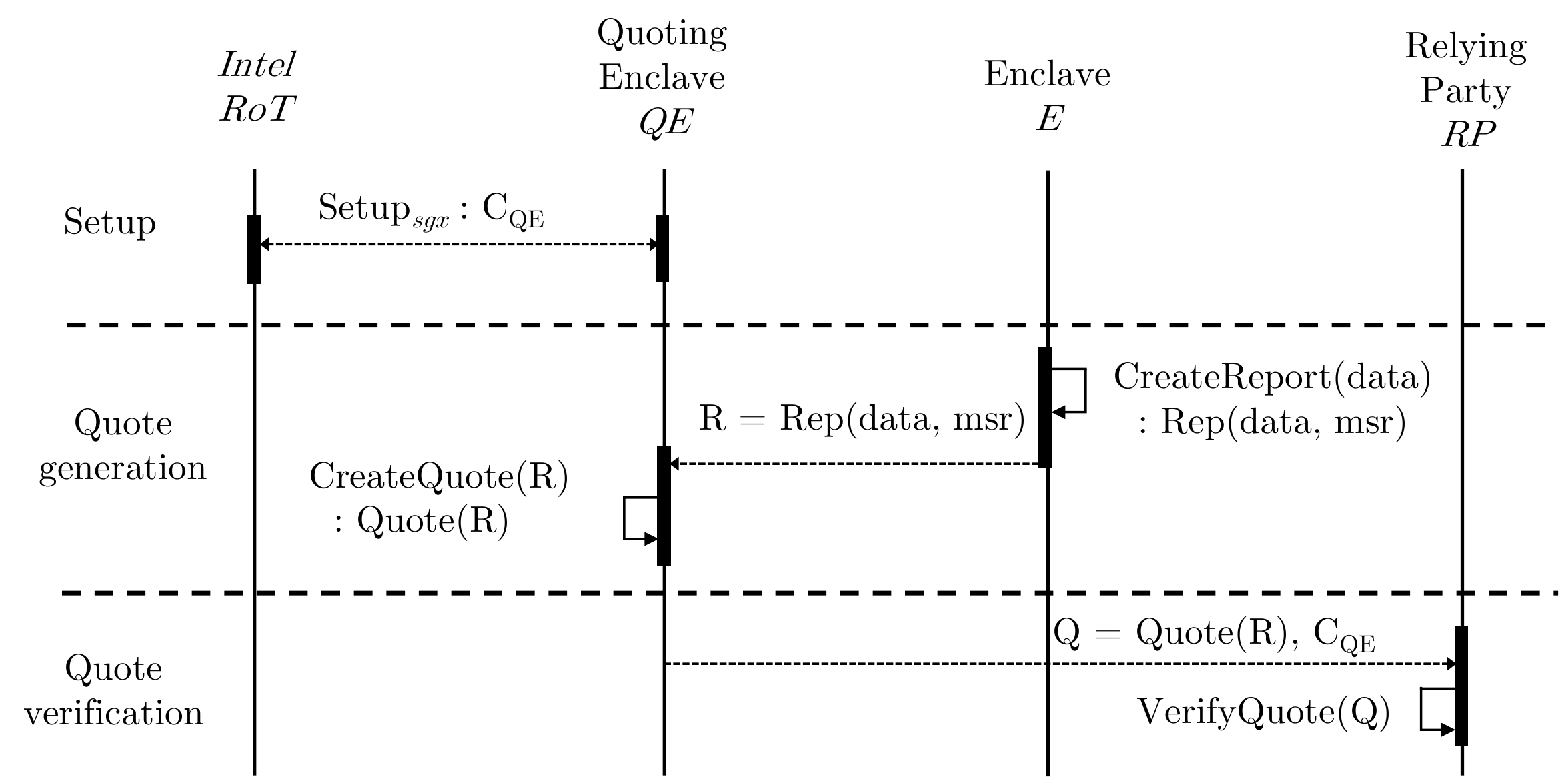}
    \caption{DCAP protocol sequence diagram.}
    \label{fig:dcap}
\end{figure*}

We split the attestation protocol into the setup, quote generation, and quote verification phases. The protocol is depicted in Figure~\ref{fig:dcap}.

% Platform setup
The platform setup phase establishes and ensures the existence of a chain of trust that extends from the Intel's root of trust to the attestation proof.
% If the root is trusted then so is the attestation proof, and one must be convinced about the validity of that proof.
% x`The root of trust is typically the manufacturer of the TEE.
During the setup phase, the SGX platform interacts with \IntelRot{}. Using a secret shared in the manufacturing process, the platform can attest itself and the quoting enclave to the root of trust. Once this attestation is successfully carried out, the root of trust certifies the quoting enclave, that is, the root of trust produces a certificate $C_{QE} = (\Qek_{pb}, ppid, asign_{IR}(\langle{}\Qek_{pb}, ppid\rangle{},\IntelRotk))$. 
% We capture this phase by the function $\text{Setup}_{SGX}(\Qek_{pb}, \IntelRotk)$ which produces $\textbf{C}$. 
We assume that this phase happens successfully as the SGX platform is being set up so that $C_{QE}$ is made publicly available.

The quote generation phase, if successfully executed, produces a quote which is a tuple ($msr, plat, data, sig)$ where $msr$ is the measurement of the enclave being attested, $plat$ is a data structure containing information about the SGX platform, $data$ is a vector of ``free" data generated by the enclave being attested, and $sig \equiv asign_{QE}(\langle{}msr,plat,data\rangle{}, \Qek)$ is the signature of the quoting enclave on these other quote elements -- the notation $\langle{}e_1,\ldots,e_n\rangle{}$ provides the ordered concatenations of elements $e_1$ through to $e_n$. It is a statement that an enclave with measurement $msr$ was running in a authentic SGX platform with characteristics given by $plat$ and it provided data $data$ when taking part in the attestation protocol. The quote is only produced if \E{} provides a \emph{local attestation report}. When the enclave with measurement $msr$ invokes the SGX instruction \textproc{EREPORT} passing $data$ as an argument, it creates such a report with which the quoting enclave can verify the integrity of $data$ and its provenance from enclave $msr$. 

Given the expected enclave measurement $msr_{exp}$, the expected data $data_{exp}$, a quote $Q = (msr, plat, data, sig_{QE})$, a certificate $C_{QE} = (\Qek_{pb}, ppid, sig_{IR})$, \RP{} can execute quote verification process consisting of: (i) verifying the signature $sig_{IR}$ using $averi_{IR}(\langle{}\Qek_{pb}, ppid\rangle{}, sig_{IR}, \IntelRotk_{pb})$, (ii) verifying $sig_{QE}$ using $averi_{QE}(\langle{}msr,plat,data\rangle{}, sig_{QE}, \Qek_{pb})$, (iii) checking that $msr$ and $data$ corresponds to the expected enclave measurement $msr_{exp}$ and $data_{exp}$. Optionally, in some usage scenarios the relying party may also verify that the $ppid$ and $plat$ match expected values or satisfy some other criteria. We use the function \textproc{VerifyQuote}($msr_{exp}, data_{exp}, Q, C$) to capture the validations (i-iii) of the quote verification phase.

% Simplified but accurate enough. PCE and platform information.
Our simplified protocol abstracts away the details and complexity of DCAP while focusing on its essential behaviour. The fully-fledged DCAP protocol relies on another architectural enclave (the Provisioning Certification Enclave) in the setup phase, and the certification of the quoting enclave is given by a certificate chain, whereas our protocol abstract that chain by a single certificate. We do not detail what is in the $plat$ structure as the goal of this paper is not to discuss the practical intricacies of an SGX platform. 
% Multiple quoting enclaves can also co-exist in an SGX platform, but this detail is immaterial to the sort of guarantee that we want to establish.

Despite its simplicity, our protocol still provides achieves the protocol's goal given the threat model and trust assumptions defined, as demonstrated by our formal analysis. Note that a quote is not \emph{directed} at a specific verifier: any relying party possessing Intel's root of trust key can verify the quote and SGX platform certificate.

\subsection{Remote attestation for SEV machines}
\label{sec:sev_attestation}

% SEV remote attestation and secret provisioning
Compared to SGX, SEV's attestation primitives are not as flexible giving rise to an attestation protocol that is arguably more restrictive and intricate. The attestation protocol takes place as the SEV guest VM is being created, and includes a provisioning step. In this paper, we are concerned with the attestation protocol and infrastructure of SEV pre-SNP. As for SGX, we propose an abstracted protocol that focus on the relevant functionality implemented by the fully-fledged SEV protocol.

The protocol involves the following parties: the AMD's secure processor of the attested platform \PSP{}, AMD's root of trust service \AmdRot{}, and the guest VM owner \GO{}, and its attested guest VM \SVM{}. \AmdRot{} is in charge of certifying the platform's \PSP{}, while \GO{} interacts with \PSP{} to attest, provision, and create \SVM{}.

\subsubsection*{Protocol goal.} 
% This protocol is used to allow \GO{} to attest \SVM{}, attest the SVM and provision a secret that is available to SVM upon launch. 
The protocol produces a \GO{}-directed attestation proof, a \emph{measurement} in SEV terminology\footnote{A SEV measurement is different from a SGX measurement. The latter refers to the digest of the enclave's code, whereas the former is a digest calculated from the VM firmware code but it also includes some platform and launch-policy information as well as a nonce biding the measurement to a particular VM launch session.} and a \emph{SEV platform certificate}, and provisions \SVM{} with \GO{}-generated secret \textit{S}. Once the protocol is completed, \GO{} is convinced of the authenticity of \SVM{}'s TCB, and that \textit{S} could only have been provisioned to \SVM{}.

\subsubsection*{Threat model and trust assumptions.} The same threat model and trust assumptions used for the SGX protocol are used in the analysis of the SEV protocol, with the exception that, here, we consider the SEV TCB and platform and AMD RoT service as trusted elements as opposed to the SGX and Intel counterparts, of course. Here, a compromised platform would allow the attacker to obtain any information that \PSP{} knows, including the cryptographic keys it manages. We do not allow SEV VM migration. We do not consider memory-remapping, rollback, or fork attacks; we assume integrity-checking mechanisms can be put in place to prevent those. Moreover, our trust in the attested SEV TCB is intended to prevent all architectural attacks --- including the ones affecting attestation primitives~\cite{Buhren19,Wilke21}. This assumption allows us to to analyse the security properties of the protocol itself, as opposed to weaknesses linked to the bad design/implementation of the underlying primitives.

\subsubsection*{Cryptographic schemes.} The protocol involves the following cryptographic schemes:
\begin{itemize}
    \item \AmdRot{} uses an asymmetric signature scheme defined by functions $agen_{AR}()$, $asign_{AR}(m,k)$, and $averi_{AR}(m,s,k_{pb})$;
    \item \AmdRot{} key pair $(\AmdRotk_{pb}, \AmdRotk)$, public and private elements, respectively, is generated using $agen()_{AR}$ and used by the root of trust to issue SEV platform certificates;
    \item \PSP{} and \GO{} rely on the asymmetric secret-negotiation scheme with key-generation function $sngen()$ and secret computation function $snsec(K_{pb},K)$, where $K_{pb}$ and $K$ are public and private key elements of the scheme. Diffie-Hellman key-sharing scheme is an instatiation of such a scheme. 
    \item \GO{} generates the key pair $(GoSn_{pb}, GoSn)$ using $sngen()$. 
    \item \PSP{} generates a key pair $(\PspSn_{pb}, \PspSn)$ using $sngen()$.
    \item \PSP{} and \GO{} rely on a key-derivation function $sder(Sd)$, where $Sd$ is a derivation seed.
    \item \PSP{} and \GO{} rely on the symmetric encryption scheme defined by key-generation function $sgen_{E}()$, encryption function $senc(m,k)$, and decryption function $sdec(m,k)$, where $m$ is a message and $k$ is a scheme's key.
    This scheme is used for encrypting key-wrapping interactions and transported messages between them.
    \item \PSP{} and \GO{} rely on the message authentication code (MAC) scheme defined by key-generation function $sgen_{I}()$, signing function $ssign(m,k)$, and verification function $sveri(c,k)$, where $m$ is a message, $c$ is an authentication code, and $k$ is a scheme's key.
    This scheme is used for integrity-protecting key-wrapping interactions and transported messages between them.
\end{itemize}

Here and in our protocol description we are relying on a single symmetric encryption scheme and a single MAC one for the sake of simplicity. However, one could use multiple schemes, one for each different application, without affecting the protocols' guarantees.

\subsubsection*{Protocol.} 
\begin{figure*}
    \centering
    \makebox[\textwidth][c]{\includegraphics[width=1.2\textwidth]{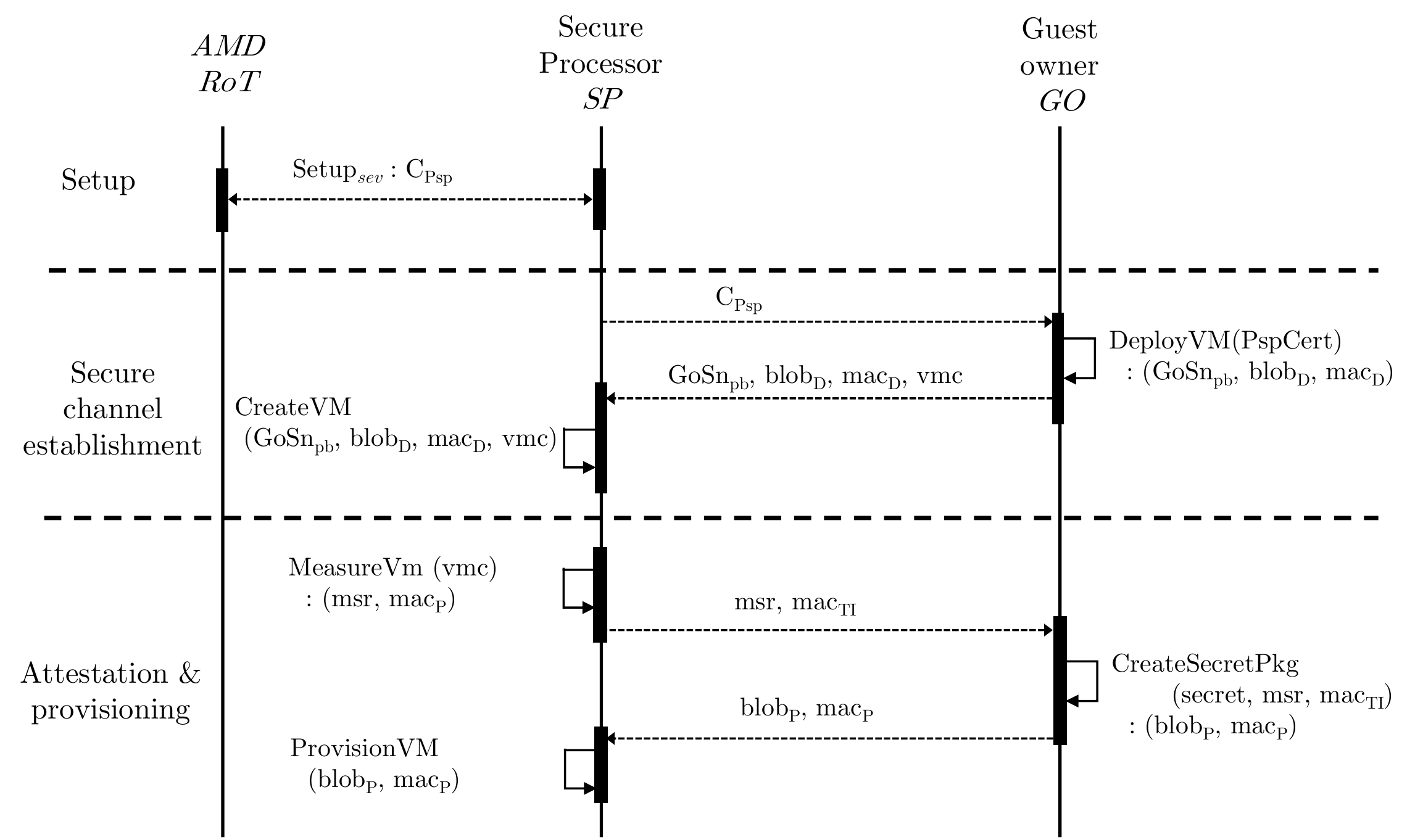}}
    \caption{SEV remote attestation protocol sequence diagram.}
    \label{fig:sevra}
\end{figure*}

We divide the protocol execution into three phases: SEV platform setup, secure-channel establishment, VM validation \& provisioning, all of which we detail next. The protocol is depicted in Figure~\ref{fig:sevra}.

% Platform setup
The platform setup phase for the SEV protocol is very similar to the one that we presented for SGX. It involves only \PSP{} and the AMD's root of trust service.
It establishes a similar chain of trust, providing similar guarantees, and it also relies on a fused pre-shared secret for platform authentication. So, when successfully executed, this phase produces the SEV platform certificate $C_{Psp} = (\PspSn_{pb}, asign_{AR}(\PspSn_{pb},{\AmdRotk}))$.
We assume that this phase is successfully completed at the time the platform is set up and that this certificate is made publicly available. Notice that, unlike SGX platform certificates, SEV certificates (by AMD's design) do not contain a platform identifier. In our protocol, we will use \PSP{}'s public key $\PspSn_{pb}$ to uniquely identify a particular SEV platform.

% Secure channel establishment: guest owner side
During the secure-channel establishment phase, \PSP{} and \GO{} interact to set up a communication channel. \GO{} obtains the PSP certificate $C_{Psp} = (\PspSn_{pb}, sig)$ for the platform and verifies it using $averi_{AR}(\PspSn_{pb}, sig, \AmdRotk_{pb})$. At this point, \GO{} generates the (shared) secret $Ss = snsec(\PspSn_{pb}, GoSn)$, which is used in turn to generate keys $Kek$ and $Kik$ via the key derivation function $sder$. These two \emph{key-wrapping keys} (as per SEV terminology) are then used to transmit the pair of freshly generated transport keys $Tek = sgen_{E}()$ and $Tik = sgen_{I}()$ generated by \GO{}. It creates the \emph{deploy package message} ($GoSn_{pb}$,$blob_{D}$, $mac_{D}$, $vmc$) to be transmitted to \PSP{} where $vmc$ is \SVM{}'s firmware code, $blob_{D} = senc(\langle{}Tek,Tik\rangle{},{Kek})$ is the encrypted-keys blob, and $mac_{D} = ssign(blob_{D},{Kik})$ its authentication code. Note that \SVM{}'s code is transmitted in the clear without any integrity protection.

% Secure channel establishment: PSP side;
Upon receiving the message ($GoSn_{pb}, blob, mac, vmc)$, \PSP{} can derive the same secret $Ss$ using $snsec(GoSn_{pb}, \PspSn)$, and use it to derive keys $Kek$ and $Kik$ by the same key derivation process as \GO{}. These keys can be, in turn, used to decrypt the received blob and recover the transport keys, i.e. $\langle{}Tek,Tik\rangle{} = sdec(blob,{Kek})$, and authenticate and integrity check them with $sveri(blob,mac,{Kik})$. Therefore, at the end of this phase, \PSP{} and \GO{} have set up a secure communication channel by sharing $Tik$ and $Tek$. 

The VM attestation \& provisioning phase proceeds as follows. \PSP{} prepares \SVM{} with code $vmc$ for launch and calculates the corresponding code digest $dig$. Then, it creates the measurement $msr = \langle{}plat_{sev},launch_{sev}, dig, nonce \rangle{}$ where $nonce$ is a freshly generated random value. Structures $plat_{sev}$ and $launch_{sev}$ abstract information related to \SVM{}'s TCB and launch policies, respectively. \PSP{} constructs the \emph{measurement package message} ($msr$, $mac_{TI}$), where $mac_{TI} = ssign(msr, Tik)$, which is transmitted to \GO{}. 

Upon receiving message ($msr, mac)$, \GO{} validates the measurement by checking $sveri(msr, sig, Tik)$ and that the measurement $msr$ elements are as expected; it includes checking $digest(msr) = dig_{exp}$, where $digest(m)$ gives the code digest element of the measurement $m$, and $dig_{exp}$ is the digest independently computed by \GO{} using $vmc$.

% Secret provisioning
If this measurement validation succeeds, \GO{} proceeds to provision \SVM{}. It generates secret $S$, and creates the encrypted blob $blob_{P} = senc(S,Tek)$, and the corresponding authentication code $mac_{P} = ssign(\langle{}blob_{P}, msr\rangle{},Tik))$. Note that $mac_{P}$ takes into account the \SVM{}'s measurement $msr$. The \emph{secret package message} $(blob_{P}, mac_{P})$ is then sent to \PSP{}. 

Upon receiving message $(blob, mac)$, \PSP{} recovers the secret by decrypting the encrypted blob $S = sdec(blob, Tek)$, and it checks $sveri(\langle{}blob, msr\rangle{}, mac, Tik)$ to verify the secret blob's authenticity and integrity, and that it is provisioning the machine with the correct $msr$. If this verification does not succeed, this provisioning step is aborted. Otherwise, \PSP{} places the secret $S$ in an encrypted page of \SVM{}'s memory. Once this step is completed, \SVM{} is allowed to start its execution.

%% Abstraction move away from details. Do not instantiate the schemes like SEV. Certificate chain abstracted by a single certificate. Platform information + SEV policy. No nonce so no replay-atteck protection.
Our protocol focuses on the essential functionality required to prove that it achieves the desired goal given the threat model and trust assumptions defined. So, we simplify and abstract away elements as long as the intended guarantees can be delivered. For instance, the fully-fledged SEV protocol relies on a certificate chain which we ``flatten" to a single platform certificate. Moreover, we abstract platform and launch details by relying on opaque structures. Our model could rely on predicates over these opaque structures to identify ``desirable" platform and launch settings. There are many implementation details related to identifying memory ranges in the messages exchanged with \PSP{}.

% Comparison with SGX attestation.
Unlike the SGX protocol, the SEV attestation (and provisioning) is directed at the guest owner, and it does not contain any SEV-VM-provided data. Hence, a relying party cannot independently and convincingly establish an authenticated channel with a SEV VM --- the guest owner alone has this capability.

\begin{figure*}
    \makebox[\textwidth][c]{\includegraphics[width=1.5\textwidth]{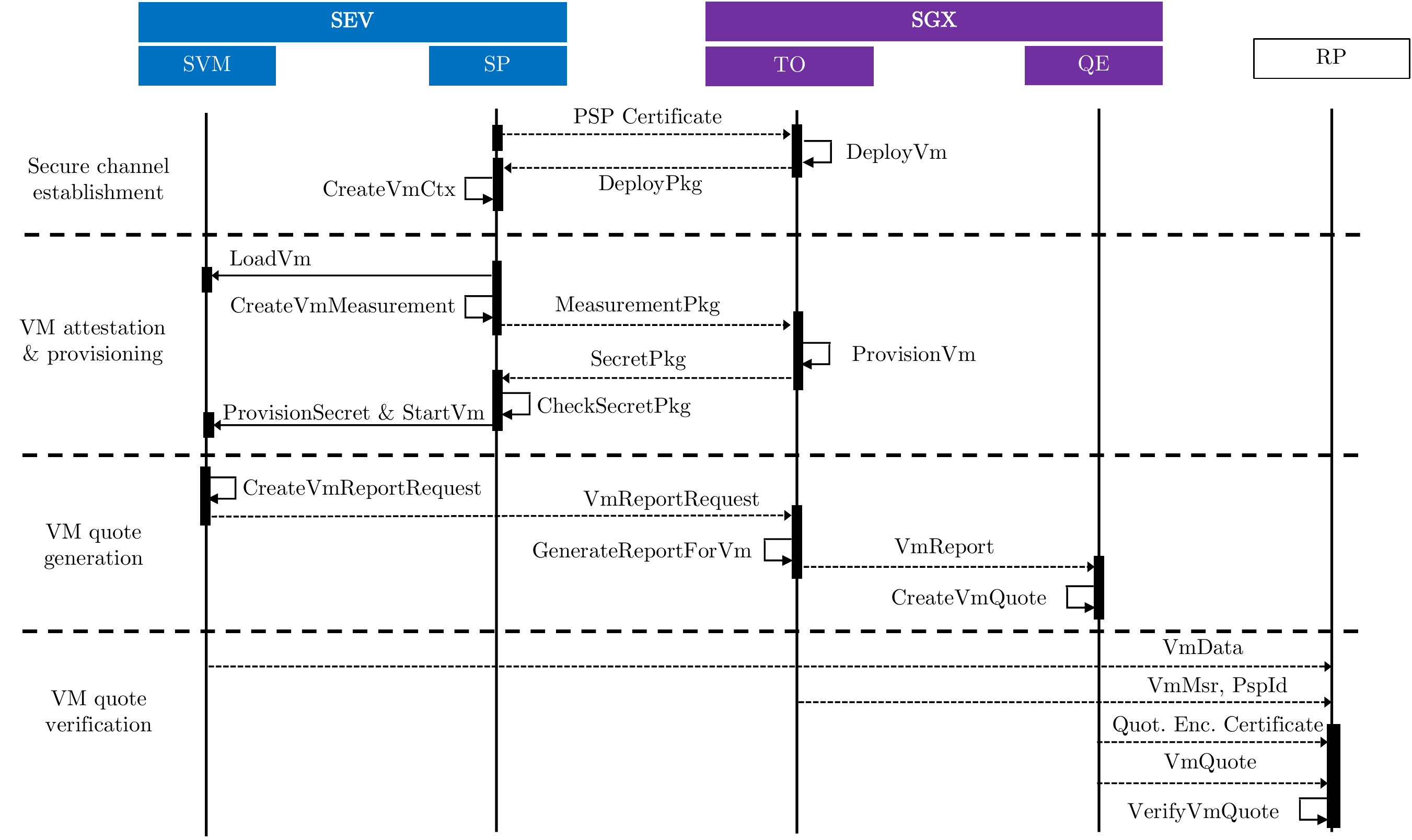}}
    \caption{Flexible SEV attestation protocol sequence diagram outline.}
    \label{fig:protocol}
\end{figure*}

\subsection{Our protocol}
\label{sec:protocol}

Our protocol is built upon the notion of a \emph{trusted guest owner}: an entity that deploys and provisions a SEV guest VM and is trusted to provide attestation reports on the deployed SEV VM's behalf. 
% Our protocol is built around an implementation of a trusted guest owner using an SGX enclave, and it combines the SGX and SEV attestation protocols that we presented.

Our protocol involves the parties in both SGX and SEV attestation protocols. However, the enclave in the SGX attestation coincides with the guest owner of the SEV attestation. So, the parties are: the trusted guest owner \TO{}, the SEV guest VM \SVM{}, the quoting enclave \QE{}, the AMD's secure processor \PSP{}, Intel's root of trust service \IntelRot{}, and AMD's root of trust service \AmdRot{}, and the relying party \RP{}.

\subsubsection*{Protocol goal.} The protocol produces an attestation proof consisting of a quote, and both SGX and SEV platform certificates. 
It authenticates both \SVM{} and \TO{}'s TCBs. The SGX platform certificate contains the Platform Provisioning ID (PPID) uniquely identifying the SGX platform instance there \TO{} was running, while the quote itself contains a digest of $\PspSn_{pb}$ --- this public key uniquely identifies the SEV platform instance where the \PSP{} and \SVM{} were running. Finally, the quote has the digest of a piece of data \textit{D} that is provided by \SVM{}. Any relying party can, then, cryptographically validate this proof and be convinced that this quote was generated using the SGX platform identified by PPID and the SEV platform identified by $\PspSn_{pb}$ with the corresponding SGX and SEV TCBs, and that \SVM{} provided \textit{D} when the protocol was executed. 

\subsubsection*{Threat model and trust assumptions.} We combine both models and assumptions of the two SGX and SEV attestation sub-protocols we use; the assumptions on \TO{} are the same as the ones made about the attested enclave \E{} in the SGX attestation protocol.
Moreover, \SVM{} is trusted not to expose the provisioned secret, which is, in our protocol, a secret key shared between \TO{} and \SVM{} - we call such a machine \emph{compliant}.

\subsubsection*{Cryptographic schemes.} We rely on the cryptographic schemes that are required by both SGX and SEV attestation protocols, which we do not restate here for the sake of brevity, plus the cryptographic hash function $hash_{TO}$ used by \TO{} in emitting reports for \SVM{}.

\subsubsection*{Protocol} We split our protocol into phases: setup, secure channel establishment, VM attestation \& provisioning, VM report generation, and verification by relying party. The protocol is depicted in Figure~\ref{fig:protocol}; we omit the setup phase from the diagram for conciseness.

The setup phase successfully carries out the setup phases of both SGX and SEV attestation protocols for the attested platforms, and it precedes the other phases of our protocol. As a result, it produces \PSP{} and \QE{} certificates $(\PspSn_{pb}, asign_{AR}(\PspSn_{pb}, \AmdRotk{}))$ and $(\Qek_{pb}, ppid, asign_{IR}(\langle{}\Qek_{pb}, ppid\rangle{},\IntelRotk))$, respectively.

\begin{algorithm}
\small
\caption{\TO{}'s code. \small We use the schemes as defined in the text, and the well-known \emph{Option} type. The enclave global variables and constants start with an uppercase letter whereas the local ones start with a lowercase one. Their types are not explicitly mentioned but they can be inferred from their usage. The constants hold the values of the corresponding public keys, and the global variables are initialised with \emph{None}. As for the types of our functions, we use \textproc{PUB}$_{x}$ to denote the public-key type of scheme identified by $x$, \textproc{SIG}$_{x}$ is a signature type, \textproc{CYP}$_{x}$ a cyphertext type, \textproc{DIG}$_{sev}$ the SEV code digest type, \textproc{MSR}$_{sev}$ the SEV measurement type, \textproc{REP}$_{sgx}$ the SGX local attestation report type, and \textproc{DAT} the VM report $data$ type.}
\label{alg:trusted_goc}
\begin{algorithmic}
\State \textbf{vars} PspId, Tik, Tek, VmDig, Msr, Cik $\gets$ None
\State \textbf{consts} \AmdRotk$_{pb}$%, \IntelRotk$_{pb}$
\State 
\Function{DeployVm}{(\PspSn$_{pb}$, sig) : PUB$_{Sn}$ $\times$ SIG$_{AR}$,  dig : DIG$_{sev}$}{ : Option(PUB$_{Sn}$ $\times$ CYP$_{kek}$ $\times$ SIG$_{kik}$)}
    \If{VmDig $=$ None $\land$ $averi_{AR}$(\PspSn$_{pb}$, sig, \AmdRotk$_{pb}$)}
        \State PspId $\gets$ Some(\PspSn$_{pb}$)
        \State VmDig $\gets$ Some(dig)
        \State (goSn$_{pb}$, goSn)  $\gets$ $sngen$()
        \State sd $\gets$ $snsec$(\PspSn$_{pub}$, goSn)
        \State kek, kik $\gets$ $sdev$($\langle{}$sd, 'sev\_kek'$\rangle{}$), $sdev$($\langle{}$sd, 'sev\_kik'$\rangle{}$)
        \State Tek, Tik $\gets$ Some($sgen$()), Some($sgen$())
        \State blob$_{D}$ $\gets$ $senc$($\langle{}$Tek, Tik$\rangle{}$, kek)
        \State mac$_{D}$ $\gets$ $ssign$(blob$_{D}$, kik)
        \State \Return Some(goSn$_{pb}$, blob$_{D}$, mac$_{D}$)
    \EndIf
    \State \Return None
\EndFunction
\State 
\Function{ProvisionVm}{msr : MSR$_{sev}$, mac : SIG$_{Tik}$}{ : Option(CYP$_{TE}$ $\times$ SIG$_{Tik}$)}
    \If{VmDig $\neq$ None $\land$ Cik $=$ None $\land$ $sveri$(msr, mac, Tik) $\land$ digest(msr) = VmDig}
        \State Msr $\gets$ Some(msr)
        \State Cik $\gets$ Some($sgen()$)
        \State blob$_{P}$ $\gets$ $senc$(Cik, Tek)
        \State mac$_{P}$ $\gets$ $ssign$($\langle{}$msr, blob$_{P}\rangle{}$, Tik)
        \State Tek, Tik $\gets$ None, None
        \State \Return Some(blob$_{P}$, mac$_{P}$)
    \EndIf
    \State \Return None
\EndFunction
\State
\Function{GenerateReportForVm}{vmdata : DAT, mac : SIG$_{CI}$}{ : Option(REP$_{sgx}$)}
  \If{Cik $\neq$ None $\land$ $sveri_{CI}$(vmdata, mac, Cik)}
    \State rpdata $\gets$ $hash_{TO}$($\langle{}$PspId, Msr, vmdata$\rangle{}$)
    \State \Return Some(EREPORT(rpdata))
 \EndIf
 \State \Return None
 \EndFunction
\end{algorithmic}
\end{algorithm}

% State machine. [DeployVM] -> [ProvisoinVM] -> [TruestedOwnerReport] -> repeat [GenerateReport for data X] -> Terminate
\TO{} plays a central role in the remaining phases of our protocol. Its code is presented in Algorithm~\ref{alg:trusted_goc}. The global variables, stored in protected memory, define the enclave's state; they are listed after the keyword \textproc{\bf vars}. The AMD root of trust public key is the only enclave constant, and is listed after keyword \textproc{\bf{}consts}. The functions describe the \emph{trusted} behaviour it can engage on. The input arguments for such a function is transmitted from unprotected to protected memory before its execution starts, output ones move in the opposite direction at the end of its execution, and its execution is confidential and integrity-protected. Note that, for a given instance of our trusted owner enclave, the implementation of our trusted functions ensures that \textproc{DeployVm} and \textproc{ProvisionVm} can only be meaningfully (without returning \textproc{None}) executed once and in this order. Function \textproc{GenerateReportForVm} can be meaningfully executed multiple times but only after the other two have meaningfully executed. We do not address the possibility of replayed calls to function \textproc{GenerateReportForVm}. For the sort of usage we envision that possibility does not seem too problematic, but we could address that in future versions of our protocol.

% How trusted owner interacts with PSP.
The secure-channel establishment and the VM attestation \& provision phases correspond to the homonyms of the SEV attestation protocol, presented in Section~\ref{sec:sev_attestation}, with \TO{} playing the part of \GO{}. 

The function \textproc{DeployVM} implements the guest owner's behaviour in this phase. Given a PSP certificate and a SEV VM code digest as input, this function carries out all the necessary certificate verification, secret negotiation, key derivations and generations on its way to create and return \TO{}'s secret-negotiation public key \textproc{goSn$_{pb}$}, the encrypted blob \textproc{blob$_{D}$}, and authentication code \textproc{mac$_{D}$} for the generated transport keys. These keys are stored in enclave global variables \textproc{Tek} and \textproc{Tik}. This function also fixes the expected code digest of the SEV VM being deployed, which is stored in the global enclave variables \textproc{VmDig}. Note that this function is only concerned with the digest of the VM code --- the code itself can be stored and communicated by untrusted components. The elements returned by this function together with the VM code itself are combined to create the deployment package message. This message is relied to \PSP{}, who carries out the rest of this phase as described in Section~\ref{sec:sev_attestation}.  

The VM attestation \& provision phase starts with \PSP{} constructing the measurement package message as per Section~\ref{sec:sev_attestation}. The function \textproc{ProvisionVM}, which implements the behaviour of the guest owner in this phase, takes as input the measurement and authentication code in that message. The function carries out the verification of the input measurement, generates a MAC-scheme key stored in \textproc{Cik}, and produces the secret encrypted blob and authentication code. The blob and code are used to create the secret package message which is sent to \PSP{}, which carries out the secret package verification and provisioning, bringing this phase to an end, as per Section~\ref{sec:sev_attestation}. The sharing of the \textproc{Cik} key via this provisioning step establishes an authenticated (but not confidential) channel between \TO{} and \SVM{}.

The VM quote generation and verification phases involves the execution of the SGX attestation protocol, presented in Section~\ref{sec:sgx_attestation}. These phases of the protocol take place after the initial three have successfully completed and \SVM{} has started. 

The VM quote generation starts with \SVM{} creating a report request $(vmdata, mac)$, where \textit{vmdata} is a piece of data generated by it, and $mac = ssign_{CI}(vmdata, \textproc{Cik})$. This report request is then communicated to \TO{} by invoking \textproc{GenerateReportForVm} with \textit{vmdata} and $mac$ as inputs. Upon successful verification of $mac$, this function creates a SGX report addressed to \QE{} containing: \TO{}'s enclave measurement $msr_{TO}$, and a digest of the public key $\PspSn_{pb}$ identifying the attested SEV platform, of \textit{vmdata} and of \SVM{}'s measurement \textit{Msr} represented as $rpdata$. This report is transmitted to \QE{} which generate the corresponding quote ($msr_{TO}$, $rpdata$, $asign_{QE}$($\langle{}$msr, data$\rangle{}$, \Qek)).

\RP{} verifies the VM quote using the function \textproc{VerifyQuote} in Section~\ref{sec:sgx_attestation}. Let $Q$ be the VM quote received, $msr_{TO}$ the enclave measurement for \TO{}, $C_{QE}$ the quoting enclave certificate, $vmdata$ the VM piece of data, $vmmsr$ the VM measurement, and $pspid$ the attested SEV platform id. \RP{} calculates the expected report data $rpdata_{exp} = hash_{TO}(\langle{}pspid,vmmsr,vmdata\rangle{})$, and checks \textproc{VerifyQuote}($msr_{TO}$, $rpdata_{exp}$, $Q$, $C_{QE}$). This validation convinces \RP{} that the protocol's goal has been achieved, namely, that the $vmdata$ was generated by a SEV VM with measurement $vmmsr$.

\subsection{Formal specification and verification}
\label{sec:verification}

To validate our proposal, we give a formal model of the flexible attestation protocol, and use the Tamarin prover to provide machine-verifiable proofs that it has the desired security properties. Hence, the protocol meets its stated goals in a setting with an unbounded number of sessions assuming a Dolev-Yao attacker and a threat model described in Section~\ref{sec:protocol}. 
We make the formal model as well as the proofs and the proof oracle needed to replicate the results publicly available at~\cite{repo}.

\subsubsection*{Protocol model}

We model the protocol by specifying all participants using multiset rewriting rules as in ~\cite{Meier2013}. Each rule is of the form $id{:}\ltr{l}{a}{r}$, where $l,a,r$ are sets of \emph{facts}. Facts in $l$ are rule \emph{premises}, facts in $r$ are \emph{conclusions} and those in $a$ are \emph{action facts} of the rule. As an example, the rule corresponding to the \textproc{DeployVm} function of the trusted owner is given in Figure~\ref{fig:rule}. First of all, the ``let'' binding only acts as syntactic sugar making the specification more readable. In the rule premises, \TO{} ensures that it is running on a initialised SGX platform (by checking the existence of a \emph{persistent} fact generated by another rule); it makes sure the PSPs certificate is already verified (by another rule); it receives the code of the guest VM \SVM{} to deploy (abstracted as a \emph{public value}); it creates a Diffie-Hellman private key as well as the transport keys. In the rule conclusions, \TO{} sends the request for guest creation and stores the necessary information in its session state. The request is created by generating the shared secret, deriving keys and encrypting/MAC-ing appropriate data. Action facts are later used to specify security properties. In addition to five protocol participants from Figure~\ref{fig:protocol} (\SVM{}, \PSP{}, \TO{}, \QE{}, \RP{}), we explicitly model Intel and AMD roots of trusts services.

\begin{figure}[t]
\small
\centering
\begin{verbatim}
rule TO_Enclave_Deploy_VM:
  let
    go_sn_pk = 'g' ^ ~go_sn
    sd = psp_sn_pk ^ ~go_sn
    kek = h(<'sev_kek', sd>)
    kik = h(<'sev_kik', sd>)
    msg_content = <'transport_keys', ~tek, ~tik>
    blob = senc(msg_content, kek)
    mac = h(<msg_content, kik>)
    deploy_package = <go_sn_pk, blob, mac, $vm_dig>
  in
  [
    !SGX_Platform_Initialied(~ppid)
    , Platform_PK_Verified(psp_sn_pk)
    , In($vm_dig)
    , Fr(~go_sn)
    , Fr(~tek)
    , Fr(~tik)
  ]--[
    TO_Enclave_Deploy_VM()
    , TO_Enclave_Secrets(psp_sn_pk, sd, kek, 
        kik, ~tek, ~tik)
  ]->[
    Out(deploy_package)
    , TO_Enclave_VM_Deployed(psp_sn_pk, ~ppid, 
        ~tek, ~tik, $vm_dig)
  ]
\end{verbatim}
\caption{One of the rewrite rules modeling the TO}
\label{fig:rule}
\end{figure}
The functional part of the formal model consists of 21 rules given in Table~\ref{tab:rules}. The rules are almost in one-to-one correspondence with the description of protocol steps given in Section~\ref{sec:protocol}. The exception are the attacker rules that we introduced to faithfully capture the threat model and allow the corruption of parts of the system. 

\subsubsection*{Attacker model}
The Dolev-Yao attacker rules are automatically embedded in the model by the Tamarin tool, but we need to add additional attacker actions to be faithful to the desired threat model. In particular, we add rules that disclose quoting enclaves and PSPs long term private keys to the attacker, corresponding to corruptions of arbitrary SGX and SEV platforms; these rules do not apply to non-compromised platforms. We also add rules to corrupt both roots of trust as a means to sanity check our model.

\begin{table}[t]
    \centering
    \resizebox{.7\columnwidth}{!}{%
    \setlength{\tabcolsep}{4pt}
    \begin{tabu}{|l|c|}
        \hline
         Rule name & Protocol party \\
         \hline
\texttt{Intel\_RoT\_Initialize} & \IntelRot{} \\
\texttt{Intel\_RoT\_Certify} & \IntelRot{} \\
\texttt{SGX\_QE\_Initialize} & \QE{} \\
\texttt{SGX\_QE\_Generate\_Quote} & \QE{} \\
\texttt{AMD\_RoT\_Initialize} & \AmdRot{} \\
\texttt{AMD\_RoT\_Certify} & \AmdRot{} \\
\texttt{SEV\_PSP\_Initialize} & \PSP{} \\
\texttt{SEV\_PSP\_Initialize\_Guest} & \PSP{} \\
\texttt{SEV\_PSP\_Launch\_Guest} & \PSP{} \\
\texttt{TO\_Enclave\_Verify\_Platform\_Cert} & \TO{} \\
\texttt{TO\_Enclave\_Deploy\_VM} & \TO{} \\
\texttt{TO\_Enclave\_Provision\_VM} & \TO{} \\
\texttt{TO\_Enclave\_Generate\_Report\_For\_VM} & \TO{} \\
\texttt{Guest\_VM\_Request\_Report} & \SVM{} \\
\texttt{RP\_Verify\_Quote} & \RP{} \\
\texttt{Compromise\_Intel\_RoT} & adversary \\
\texttt{Compromise\_SGX\_QE} & adversary \\
\texttt{Adversary\_Request\_Quote} & adversary \\
\texttt{Compromise\_AMD\_RoT} & adversary \\
\texttt{Compromise\_SEV\_PSP} & adversary \\
\texttt{Adversary\_Extract\_SEV\_Secret} & adversary \\
         \hline
    \end{tabu}}
    \caption{All the rules in the formal model.}
    \label{tab:rules}
\end{table}

We list and discuss the attacker rules related to SEV here, the rules related to SGX are similar. 
The \texttt{Compromise\_AMD\_RoT} allows the adversary to compromise the \AmdRot{} and extract the \AmdRotk{} private key. This rule was added purely for sanity checking purposes and, indeed, the main results and well as the lemmas related to SEV are falsified unless we assume the adversary did not use this rule.

{
\small
\begin{verbatim}
rule Compromise_AMD_RoT:
  [
    !AMD_RoT_Ltk(~amd_rot_ltk)
  ]--[
    Compromise_AMD_RoT()
  ]->[
    Out(~amd_rot_ltk)
  ]
\end{verbatim}
}

The \texttt{Compromise\_SEV\_PSP} allows the adversary to compromise one specific \PSP{} and extract the \PspSn{} private key of that platform. This rule models platform compromise (e.g., by side-channel attacks). We show that the main results hold even if the adversary can compromise arbitrary platforms, as long as the \emph{specific} \PSP{} used in the protocol execution is not compromised.

{
\small
\begin{verbatim}
rule Compromise_SEV_PSP:
  [
    !PSP_Ltk(~cpu_id, ~psp_sn)
    , !PSP_Pk(~cpu_id, psp_pk)
  ]--[
    Compromise_SEV_PSP(psp_pk)
  ]->[
    Out(~psp_sn)
  ]
\end{verbatim}
}

One of the modeling challenges was formalising the relationship between a measurement and the behaviour of the measured code. Using SGX as an example, we need to be able to combine the fact that the quoting enclave produced a quote with measurement $msr_E$ and data $data_E$ with the fact that measurement $msr_E$ corresponds to specific enclave code $E$ with certain behaviour when executed on trusted hardware (e.g., $E$ only provides attestation reports in which $data_E$ is in a specific format). To address this challenge in general, the framework has to support higher-order reasoning about the building blocks of protocol specification --- e.g., we need to use those building blocks both as programs that can be executed and as data that can be hashed or send over the network (perhaps to be executed on the other end). To the best of our knowledge, no protocol verification framework currently allows reasoning about such constructions. 

As our scope in this paper is limited to modeling and verifying the proposed protocol, we overcome this challenge by using a simple over-approximation of the attacker's capabilities. In the SGX setting, we assign a fixed measurement $const_{TO}$ to enclave \TO{} is running. Furthermore, we allow the attacker to obtain valid quotes with arbitrary data for any measurement \emph{except} for $const_{TO}$. Hence, we hardcode the relationship \TO{} and the measurement of its enclave in our model, and assume enclaves corresponding to all other measurements are under the control of the attacker. We take a similar approach with SEV --- we hardcode the launch digest $const_{SVM}$ of our guest VM and allow the attacker to extract secrets provisioned by the PSP from any SEV VM whose launch digest is \emph{different} from $const_{SVM}$. 
We list the rule and give more details for SEV here.

The \texttt{Adversary\_Extract\_SEV\_Secret} allows the adversary to extract a provision secret from a VM running on arbitrary \PSP{}. This rule models the fact that adversary can launch and control arbitrary VMs on an arbitrary \PSP{}. The only thing we disallow (via the \texttt{Neq} \emph{restriction}) is that the adversary extracts the secret from our specific \SVM{} whose digest a constant $const_{SVM}$ (a string \texttt{burrito\_guest\_vm} in the Tamarin model).

{
\small
\begin{verbatim}
rule Adversary_Extract_SEV_Secret:
  [
    !SEV_PSP_Guest_Running(~cpu_id, psp_sn_pk, 
        $vm_dig, ~guest_secret)
  ]--[
    Neq($vm_dig, 'burrito_guest_vm')
    , Adversary_Extract_SEV_Secret($vm_dig,
        ~guest_secret)
  ]->[
    Out(~guest_secret)
  ]
\end{verbatim}
}

\subsubsection*{Security properties and proofs} The main security property we are interested in verifying is the authenticity and integrity of the resulting VM quotes. As helper lemmas, but also as results of their own merit, we verify the security properties of both SGX attestation and SEV secure guest deployment as used in our system. The most important verified properties are informally described next, and they are followed by the corresponding Tamarin lemmas.
\begin{description}
\item[SGX quote authenticity] If \RP{} verifies a SGX quote with the measurement $const_{TO}$, with a certificate identifying the $ppid$ SGX platform, and quote data $rpdata$, then \TO{} has executed \textproc{GenerateReportForVm} function on a SGX platform identified by $ppid$ and $rpdata$ is equal to $hash_{TO}$($\langle{}$PspId, Msr, vmdata$\rangle{}$) for some PspId, Msr and vmdata. The claim holds unless the attacker has compromised the Intel root of trust or \QE{}, the quoting enclave running on platform $ppid$.
\item[Secrecy of SEV guest secrets] If \TO{} executes \textproc{ProvisionVm} with the $const_{SVM}$ parameter and a specific PspId value, then the secret being provisioned \textproc{Cik} is never known to the attacker. The claim holds unless the attacker has compromised \AmdRot{} or \PSP{}, the specific PSP whose public key is PspId.
\item[VM quote authenticity] If \RP{} verifies an SGX quote with the measurement $const_{TO}$, with a certificate identifying the $ppid$ SGX platform, and quote data that is equal to $hash_{TO}$($\langle{}$PspId, $Msr$, vmdata$\rangle{}$) for some PspId and vmdata, and the digest in measurement $Msr$ being $const_{SVM}$, then SEV VM has executed \textproc{GenerateReportForVm} while running on a SEV platform identified by PspId with the data in the request equal to vmdata. The claim holds unless one of the following is true: the attacker has compromised the Intel root of trust; the attacker has compromised \QE{}, i.e., the specific QE corresponding to platform $ppid$; the attacker has compromised the AMD root of trust; the attacker has compromised \PSP{}, i.e., the specific PSP whose public key is PspId.
\end{description}

% TAMARIN CORRESPONDANCE
We present formal statements of the main results as well as the most important auxiliary lemmas in Tamarin notation. This notation is somewhat different compared to the informal statements above so we give clarifications when needed.

In the \textbf{SGX quote authenticity} lemma below, the informal statements ``\RP{} verifies a SGX quote'' and ``\TO{} has executed \textproc{GenerateReportForVm} function'' are modelled as Tamarin \emph{action facts} (respectively, \texttt{RP\_Verify\_Quote} and 
\texttt{TO\_Enclave\_Generate\_Report\_For\_VM}). These action facts hold at timestamps when the corresponding rules are executed. The variables \texttt{ppid} and \texttt{rd} correspond to $ppid$ and $rpdata$ in the informal statement, while \texttt{k}, \texttt{d} and \texttt{v} correspond to the report hash payload --- $PspId$, $Msr$ and $vmdata$. Note that these are untyped in the lemma statement below and are, hence, quantified over all possible messages. Variables \texttt{\#i} and \texttt{\#j} are typed as timestamps. The constant SGX measurement of the \TO{} is simply a string \texttt{burrito\_enclave\_sgx\_measurement}.

{
\small
\begin{verbatim}
lemma lm_sgx_quote_authenticity:
  "All ppid #i rd. 
    RP_Verify_Quote(<'sgx_quote',
       'burrito_enclave_sgx_measurement', ppid, 
            rd>) @ i
    ==>
    (
      (Ex v d k #j. rd = h(<'report_data', k, d, 
        v>) & 
          TO_Enclave_Generate_Report_For_VM(ppid, 
            k, d, v) @ j )
      | (Ex #j. Compromise_Intel_RoT() @ j )
      | (Ex #j. Compromise_SGX_QE(ppid) @ j )
    )
  "
\end{verbatim}
}

In the \textbf{Secrecy of SEV guest secrets} lemma below, the constant launch digest of the \SVM{} simply the string \texttt{burrito\_guest\_vm}. The action fact \texttt{KU} models the attacker knowledge, while \texttt{s} is the secret being provisioned to the \SVM{}.

{
\small
\begin{verbatim}
lemma lm_sev_guest_secret_secrecy:
  "All k s #i. 
    TO_Enclave_Provision_VM(k, s, 
        'burrito_guest_vm'
    ) @ i 
    ==> 
    (
      (not Ex #j. KU(s) @ j)
      | (Ex #j. Compromise_AMD_RoT() @ j )
      | (Ex #j. Compromise_SEV_PSP(k) @ j )
    )
  " 
\end{verbatim}
}

In the \textbf{VM quote authenticity} lemma below, notation is similar same as in the previous two lemmas. Note that we do not include the platform and the policy metadata $plat\_sev$ and $launch\_sev$ to the SEV measurement as they do not play a security-related role on the level of abstraction used on our model. Instead, the SEV measurement is just a pair consisting of a nonce (modelled by variable \texttt{m}) and the launch digest of the \SVM{}.

{
\small
\begin{verbatim}
lemma lm_burrito_quote_integrity_strong:
  "All ppid d k m #i. 
    RP_Verify_Quote(<'sgx_quote', 
      'burrito_enclave_sgx_measurement', ppid, 
          h(<'report_data', k, <m, 
            'burrito_guest_vm'>, d>)>
    ) @ i
    ==>
    (
      (
        Ex ts #j. 
          d = <'burrito_report', ts>
          & Guest_VM_Request_Report(k, ts) @ j
      )
      | (Ex #j. Compromise_Intel_RoT() @ j )
      | (Ex #j. Compromise_SGX_QE(ppid) @ j )
      | (Ex #j. Compromise_AMD_RoT() @ j )
      | (Ex #j. Compromise_SEV_PSP(k) @ j )
    )
  "
\end{verbatim}
}

We prove all results using the Tamarin prover's automated procedure with a custom proof oracle that was necessary to achieve proof termination. In addition to the main results stated above, we prove weaker variants of the claims above where we disallow the attacker from compromising any SGX or SEV platform. We also prove a number of helper lemmas and a number of sanity-checking lemmas in order to test the model itself. Most notably, we show that all the premises for main lemmas are indeed necessary by demonstrating the existence of an attack when any of the premises is removed.

\subsection{Implementation and Evaluation}

To demonstrate how our protocol works in practice, we have created an implementation of our trusted guest owner, which can be applied to any compliant SEV VM --- we have published our code~\cite{repo}. Our prototype relies on (i.e., instantiate the abstract SEV and SGX protocols we present with) the fully-fledged versions of the SEV pre-SNP and SGX DCAP attestation protocols.

Our trusted owner enclave implementation uses the SGX SDK~\cite{SGXSDK} to capture the behaviour described in Algorithm~\ref{alg:trusted_goc}. The SGX SDK provides two main abstractions for the development of enclaves: trusted functions, which are called \textit{ecalls}, and untrusted ones, which are called \textit{ocalls}. The enclave functions are described by ecalls, which can, in turn, rely on ocalls to execute untrusted privileged code. Our functions \textproc{DeployVm}, \textproc{ProvisionVm}, and \textproc{GenerateReportForVm} are all implemented as ecalls, and they take into account the fully-fledged SEV attestation operations and data formats. So, for instance, \textproc{DeployVm} checks the SEV certificate chain to authenticate the secret negotiation key as opposed to our single certificate abstraction. Our implementation uses the code of the \textit{SEV-Tool}~\cite{sevtool} as a library to carry out a number of operations related to the SEV attestation protocol --- this standalone tool has been created to help developers operate SEV VMs and platforms.

As a proof of concept and to evaluate how our protocol fares in practice, we applied it to the generation (i.e. training) of machine learning (ML) models. We use our protocol as a way to create a notion of \emph{model accountability} in the sense that VM quotes can link a specific model with the training algorithm and data set that was used to create it. This sort of quote could be used, for instance, in the context of regulated ML, where one could \emph{a posteriori} be interested in analysing if a model was created in an unbiased/fair way. 

The SEV VM that we create runs a single service, called \emph{tf\_service}, at startup and shuts itself down after the service execution has finished. This service executes (via a Docker container) a Tensorflow~\cite{tensorflow15} script that creates a ML model, export into file \verb|model.tar.gz|, and we capture the standard output of this script into file \verb|stdout|. After creating these files, it produces a VM quote containing a hash of these two files as the VM quote report data. Thus, a relying party can verify that a given model was generated with a given data set and script. Note that the data could even be kept private up until the point it needs to be divulged to a regulator/auditor to ensure the appropriate generation of the associated model.

\begin{table}[H]
    \centering
    \resizebox{\columnwidth}{!}{%
    \setlength{\tabcolsep}{4pt}
    \begin{tabu}{|l | c | c | c | c | c |}
        \hline
         Name & Deploy & Provision & GenReport & VmLife & Over. (\%) \\
         \hline
         advanced.py & 0.118 & 0.088 & 0.139 & 198.268 & 0.174 \\
         bidirectional.py & 0.121 & 0.0911 & 0.132 & 532.250 & 0.065 \\
         knowledge.py & 0.122 & 0.103 & 0.132 & 1140.456 & 0.031 \\
         beginner.py & 0.128 & 0.087 & 0.123 & 92.217 & 0.367 \\
         text.py & 0.118 & 0.089 & 0.134 & 98.184 & 0.347 \\
          % \rowfont{\color{blue}}
         text\_trans.py & 0.129 & 0.081 & 0.130 & 1648.881 & 0.021 \\
         % \rowfont{\color{blue}}
         cnn.py & 0.122 & 0.089 & 0.135 & 339.739 & 0.101 \\
         % \rowfont{\color{blue}}
         keras.py & 0.121 & 0.094 & 0.134 & 117.956 & 0.295 \\
         % \rowfont{\color{blue}}
         preprocessing.py & 0.120 & 0.101 & 0.136 & 98.965 & 0.360 \\
         % \rowfont{\color{blue}}
         classification.py & 0.114 & 0.097 & 0.143 & 889.031 & 0.039 \\
         % \rowfont{\color{blue}}
         imbalanced.py & 0.118 & 0.086 & 0.149 & 310.548 & 0.113 \\
         % \rowfont{\color{blue}}
         word2vec.py & 0.116 & 0.093 & 0.131 & 117.545 & 0.289 \\
         \hline
    \end{tabu}}
    \caption{Accountable ML evaluation results.}
    \label{tab:evaluation}
\end{table}

Our VM is based upon the Alpine~\footnote{https://www.alpinelinux.org/} Linux distribution. It is relies on a modified SEV-ready kernel, an initial ramdisk that includes a root filesystem (containing the \emph{tf\_service} and its dependencies), and a fixed kernel command line --- these are the elements necessary to boot a Linux VM. The hashes of these three pieces of information are recorded in the initial VM firmware and are, hence, part of the VM measurement that can be verified by the relying party. The root filesystem is setup in main memory as opposed to disk. 

We point out that our machine \emph{does not} rely on the typical attestation scenario that is suggested by AMD, i.e. using a guest-owner-encrypted disk for which the key is provisioned using the SEV attestation protocol. Of course, once a VM has been setup using our protocol (and an initial root filesystem in main memory like we do), it could include a routine to create an encrypted disk whose key would remain protected in main memory. So, our protocol and example VM could still accommodate disk encryption seamlessly.

Our evaluation takes into account 12 Tensorflow scripts. For each of them, we create corresponding VMs as explained and carry out deployment, provisioning, and report generation using our trusted owner, as per our protocol. The results of executing these VMs is presented in Table~\ref{tab:evaluation}. We use a AMD machine using an EPYC 7402P 24-Core processor to run the VMs and an Intel machine with a Intel(R) Xeon(R) E-2288G CPU @ 3.70GHz processor. In this evaluation, we measure the times taken to perform each of the trusted owner functions --- they include network latency as we use a remote trusted owner. The overhead is calculated as the (Deploy+Provision+GenReport)*100/VmLife; it gives the percentage of time taken by the trusted owner operations with respect to the entire VM execution (VmLife).

As expected, the timings for executing trusted owner operations are fairly constant and independent of the VM lifetime (and execution complexity). Note that trusted owner operations are of fixed type and size so those are independent of the type of the VM being run. Moreover, the overhead imposed by our protocol is minimal: in all cases it came under 0.5\% of the VM execution time. Therefore, unsurprisingly, our protocol delivers its guarantees without incurring in significant VM-execution overheads.

\subsection{Discussion}
%% The problem can be extended to that of a trusted deployer for an intricate platform. In that case, the trusted deployer would instantiate a number of services and its report would account for the entire infrastructure.

Our protocol can be extended to accommodate a more generic and ambitious application. Instead of a single SEV VM, we could use the same principles to create a \emph{trusted deployer} that sets up and attests an entire \emph{trusted (and possibly heterogeneous) infrastructure}. Instead of having to attest the components of that infrastructure individually, possible using different protocols with varied levels of flexibility depending on the heterogeneity of the trusted components, the extended version of our protocol would allow a trusted deployer, with a flexible attestation mechanism and the capacity to deploy all the other components, to generate a single attestation report on the infrastructure's behalf. A relying party would, therefore, enjoy a simple and flexible protocol to attest the infrastructure.

Our work creates and promotes a new line of research, namely, exploring \emph{synergies between TEE implementations}. SGX provides a flexible and simple attestation mechanism and, arguably, subpar application portability, whereas SEV pre-SNP offers application portability and a overly-rigid attestation protocol. Our protocol confers SGX-like attestation to a SEV VM, thereby bringing out the best combination of application portability and attestation flexibility. Intel and AMD have, recently, proposed TEE architectures and implementations, in the form of SEV SNP~\cite{sevsnp20} and TDX~\cite{TDX}, that offer both of these qualities. However, these architectures are still immature in comparison to SGX and (pre-SNP) SEV. At the time of writing (May 2023), there hardware supporting  TDX and not generally available, software support for SEV SNP is immature, and no could providers expose the flexible attestation intreface of SEV SNP. To illustrate more concretely the lack of maturity of SEV SNP as of now, the AMD-designed SEV software stack disables the VM firmware recording of kernel, initial ramdisk, and kernal command line measurements\footnote{\url{https://github.com/AMDESE/qemu/blob/3b6a2b6b7466f6dea53243900b7516c3f29027b7/target/i386/sev.c\#L1830}}. The current absence of this feature prevents the sort of attested boot that is so useful in establishing a chain of trust on a SEV VM; we use, for instance, this attested boot in our implementation. As for TDX, inconsistencies have been outlined~\cite{Sardar21,Sardar23} on the specifications proposed by Intel,\footnote{\url{https://www.intel.com/content/www/us/en/developer/articles/technical/intel-trust-domain-extensions.html}} illustrating even its theoretical immaturity. Our protocol could be adapted to use SEV SNP or TDX as the technologies behind the guest VMs; in the context of a heterogeneous infrastructure, for example. Thus, our protocol offer similar guarantees predicated on the trustworthiness of more mature TEE implementations. In any case, our work demonstrates the validity of this type of research by proposing an example of such a synergistic TEE combination. Moreover, even when these new technologies become mature, our protocol will still be relevant as it will provide application portability and attestation flexibility for platforms that support SEV pre-SNP but do not support SEV SNP or TDX.

We could also extend our protocol in different practical ways to allow the trusted owner and SEV VM to exchange other types of information. Our protocol creates an authenticated channel between trusted owner and SEV VM by sharing a shared MAC key. We could extend our protocol to create an authenticated \emph{and confidential} channel between them by passing additionally a shared encryption key. The SEV VM and trusted owner could also have their APIs extended to exchange other pieces of verifiable information. For instance, they could both offer a remote function to provide a verifiable hardware-generated random string of bits. They could combine this string with a locally generated one to create a random ``stronger" source of randomness.

%% Limitations: SEV pre-SNP does not guarantee integrity only confidentiality. Justify as an orthogonal problem. Relying now on two infrastructures, if one is flawed, the protocol guarantees cannot be enforced. Possibility of secure channel?
Our protocol and implementation has also some limitations. A flaw in either of the TEE implementations that we rely upon can thwart the guarantees/goals of our protocol, as we assume both SGX and SEV TCBs to be trusted. That limitation is inherent to any combination of TEE implementations that makes this assumption. Moreover, in terms of our implementation, the SEV version that we use does not offer integrity protection; only SEV SNP gives integrity guarantees. We could implement our protocol using any SEV-like TEE implementation, with or without integrity protection, provided that the required attestation primitives are available.

\section{Related Work}
\label{sec:related_work}

In this section, we examine papers that focus on hardware-based TEEs and remote attestation protocols involving them. 

A number of applications and extensions to the SGX attestation protocols have been proposed. From incorporating attestation information into the TLS protocol~\cite{Knauth18}, to proposing flexible attestation verification infrastructures~\cite{Chen19}, to proposing flexible mutual attestation protocols~\cite{Chen22}. Kucab \textit{et al.}~\cite{Kucab21} propose a protocol that involves similar parties but is very different in many ways to ours. They use SGX attestation to perform an integrity check on the filesystem of (non-SEV) VMs at startup.

Another line of research consists of identifying vulnerabilities and attacks specifically targeting attestation primitives~\cite{Swami17,Buhren19,Wilke21}. Swami has shown that some of the privacy guarantees are thwarted by Intel's EPID design~\cite{Swami17}. Buhren \textit{et al.}~\cite{Buhren19} has shown how the PSP firmware can be updated to a version that allows the extraction of the cryptographic keys managed by the PSP. Wilke \textit{et al.}~\cite{Wilke21} have shown how the memory-permutation insensitivity of the SEV launch measurement can be exploited in a way that allows the VM to execute arbitrary code and yet its original launch measurement remains unchanged. We regard these works as complementary to ours. The findings about SGX's EPID can improve its privacy guarantees, and as a consequence, the benefits it could bring if it was used as part of our protocol. The other two SEV attacks are prevented by our protocol assumptions requiring the attested SEV TCB to be trusted and platform to not be compromised; we focus on the analysis of the cryptographic protocol itself by assuming that the underlying primitives are trusted. These papers provide, then, guidelines to harden attestation primitives so that our assumptions are validated and our protocol can deliver on its guarantees.

Studies have compared TEE implementations and their attestation protocols~\cite{Menetrey22,Mofrad18,Niemi22,Gottel18}. They limit themselves to point out the different characteristics of such protocols without identifying and exploring interesting synergies like we do.

Some papers have used formal techniques to describe and analyse attestation protocols involving trusted hardware. For instance, the Direct Anonymous Attestation scheme, proposed as an attestation mechanism to Trusted Platform Modules (TPMs), has been formally described~\cite{Brickell04}, and analysed using Tamarin~\cite{Wesemeyer20}. SGX's EPID, DCAP, and TDX attestation mechanisms have been formaly analysed using ProVerif~\cite{Sardar20,Sardar20b,Sardar21}. While these works focus on the detailed/concrete version of SGX's schemes, our protocol and formalisation is based upon an abstract and minimalist SGX scheme as our focus is on the interplay of SGX and SEV attestation as opposed to any of those individually. Hence, there is a degree of overlap between our work and theirs, but there is also a degree of complementarity: showing that concrete versions of these protocols achieve the desired goals demonstrate that we can instantiate our abstract SGX-like subprotocol with a concrete instance and achieve the goals and guarantees of our protocol as we expected. Arfaoui \textit{et al.} have proposed a new scheme to remotely attest a hypervisor and its (non-SEV) VMs, with a formal proof of their \emph{authorized linked attestation} protocol~\cite{Arfaoui22}. Their protocol design, trust assumptions, threat model, and protocol goals are completely different from ours.

We have found only another work that combines different TEE architectures. Zhao \textit{et al.}~\cite{Zhao22} propose a framework, called \emph{vSGX}, by which one can emulate the behaviour of a SGX enclaves inside a SEV VM. The main purpose of that work is to allow unmodified SGX enclave binaries to run on SEV hardware. Thus, they do not combine TEE implementations like we do, but they implement the execution model specific to a TEE architecture on top of another. The scheme that they propose for remote attestation relies on a \emph{provider} to provision vSGX enclaves with ``fused secrets." Note that secretly providing this ``fused secret'' requires the \textit{directed}, rigid SEV remote attestation. That framework could move away from such a \emph{directed} and provider-centric attestation scheme to a more flexible one by employing our protocol to carry the remote attestation of their virtual enclaves.

Many papers have analysed TEE implementations more generally~\cite{Hetzelt17,Costan16,Subramanyan17,Pinto19,Schneider22}, and a considerable number of works have identified vulnerabilities and attacks on SEV~\cite{Morbitzer18,Wilke20,Morbitzer21,Li19,Li21,Morbitzer19,Werner19,Radev20} and SGX~\cite{Bulck18b,Biondo18,Bulck18,Schaik20,Canella19,Bulck20,Chen19b}. These papers provide either: insight to designers of TEEs so that they can improve them so their platforms are more secure, guidelines to TEE operators so that they can put in place appropriate mitigation strategies to ensure their TCBs can be trusted. So, they are, arguably, complementary to ours in the sense that they help establish in practice the assumptions that we make in formalising and analysing our protocol.

\section{Conclusion}
\label{sec:conclusion}

We propose a cryptographic protocol that explores a synergy between SGX and SEV: it brings together the flexibility of SGX's remote attestation to the application portability of SEV --- neither of these two TEE implementations offer this combination of features independently. Our protocol relies on the notion of a \emph{trusted guest owner}, implemented in an SGX enclave, that is in charge of deploying, attesting, and provisioning a SEV VM. The latter can rely on the former to generate attestation reports on its behalf. 
Moreover, we formally demonstrate that our protocol enforces security properties related to the authenticity of quotes and confidentiality of provisioned secrets using Tamarin. Furthermore, we demonstrate with an application to machine-learning-models accountability how it can be used in practice while incurring negligible overheads.

We plan to further explore the extensions to our protocol that are required to apply it to the remote attestation of an infrastructure of heterogeneous trusted components.

\bibliographystyle{plain}
\bibliography{reference}

\end{document}